\documentclass{nature}

\title{Noisy defects in a doped Mott insulator}

\author{F. Massee$^{1,*}$, Y. K. Huang$^2$, M. S. Golden$^{2}$ \& M. Aprili$^1$}

\usepackage{lineno}
\usepackage[english]{babel}
\usepackage{amssymb,amsmath}
\usepackage{graphicx}
\usepackage{multirow}
\usepackage{bigstrut}

\begin{document}

\maketitle

\begin{affiliations}
 \item Laboratoire de Physique des Solides (CNRS UMR 8502), B\^{a}timent 510, Universit\'{e} Paris-Sud/Universit\'{e} Paris-Saclay, 91405 Orsay, France
 \item Institute of Physics, University of Amsterdam, 1098XH Amsterdam, The Netherlands
\end{affiliations}

\begin{abstract}
Detailed understanding of the role of single dopant atoms in host materials has been crucial for the continuing miniaturization in the semiconductor industry as local charging and trapping of electrons can completely change the behaviour of a device. Similarly, as dopants can turn a Mott insulator into a high temperature superconductor, their electronic behaviour at the atomic scale is of much interest. Due to limited time resolution of conventional scanning tunnelling microscopes, most atomic scale studies in these systems focussed on the time averaged effect of dopants on the electronic structure. Here, by using atomic scale shot-noise measurements in the doped Mott insulator Bi$_{2}$Sr$_{2}$CaCu$_{2}$O$_{8+x}$, we visualize sub-nanometer sized objects where remarkable dynamics leads to an enhancement of the tunnelling current noise by at least an order of magnitude. From the position, current and energy dependence we argue that these defects are oxygen dopant atoms that were unaccounted for in previous scanning probe studies, whose local environment leads to charge dynamics that strongly affect the tunnelling mechanism. The unconventional behaviour of these dopants opens up the possibility to dynamically control doping at the atomic scale, enabling the direct visualization of the effect of local charging on e.g. high T$_{\text{c}}$ superconductivity. 
\end{abstract}

Charging effects by defects and impurities have long been recognised as the leading cause for 1/f noise in conducting devices\cite{uren_apl_1985}. As the miniaturization of devices requires complete understanding and control of defects and trapping sites, much effort has been put into uncovering the properties of individual defects and their charging behaviour in semiconductors by using local probes such as scanning tunnelling microscopy (STM). Charge dynamics of nanometre sized junctions with e.g. molecules, nano-crystals and semiconductor quantum dots have been extensively studied \cite{chen_science_1999, banin_nature_1999, pradhan_prl_2005, loth_science_2012, schaffert_naturematerials_2013, lit_naturecommunications_2013, wickenburg_naturecommunications_2016}, as well as charging at atomic scale sites \cite{repp_science_2004, lee_nanolett_2004, teichmann_prl_2008, marczinowski_prb_2008, rashidi_naturecommunications_2016}. Thus far, the study of charged defects and their dynamics has mostly concentrated on semiconductor materials. However, many correlated electron systems show a rich phase diagram as a function of impurity doping. Doped Mott insulator materials, such as the high temperature superconducting cuprates and related ruthenates, are particularly interesting materials in this regard as the dopant atoms often reside in otherwise insulating layers - a promising environment for charge dynamics to occur. Locally detecting, and possibly manipulating, these sites would open up a new avenue to study the effect of impurities on the physical properties of the system. Thus far, however, to our knowledge no reports have been made of local dynamic charging effects in any of these systems, which seems to indicate that despite the insulating nature of the layers where the dopant atoms reside, the coupling of the dopants to the continuum conduction or valence bands is still too strong for charging to affect tunnelling on the millisecond time-scales that can typically be addressed with an STM. 

Real-time detection of charging and de-charging at time-scales on the order of the single electron tunnelling rate ($\tau$), which for typical currents ($I$) of a few hundred pico ampere is in the nanosecond range ($\tau = e/I$), is complicated as cable and stray capacitances limit the bandwidth of conventional STMs. A measurement of fluctuations in the current due to the discreteness of the electron charge, or shot-noise, on the other hand, is directly sensitive to changes in the tunnelling dynamics due to local charging effects - even if these are on the time-scale of the tunnelling process. This is because depending on the exact process, local charging can lead to ordering or bunching of electron tunnelling. This results in a reduction (F $<$ 1) or enhancement (F $>$ 1), respectively, of the current noise, which for random (i.e. Poissonian, F = 1) tunnelling is given by $S_{I} = 2eIF$, where $S_{I}$ is the shot-noise power spectral density, $e$ the electron charge, $I$ the current and $F$ the Fano factor \cite{blanter_physicsreports_2000, blanter_2010}. In order to measure shot-noise at the atomic scale on correlated electron systems, whose often weakly van der Waals bound layered structures generally require high junction resistances, circuitry operating in the MHz regime has recently been developed \cite{massee_arxiv_2018, bastiaans_arxiv_2018}. Thus equipped we set out to look for signatures of charging at atomic scale defects on time-scales on the order of the tunnelling process in the near-optimally doped high temperature superconductor Bi$_{2}$Sr$_{2}$CaCu$_{2}$O$_{8+x}$ (Bi2212).

\begin{figure}
	\centering
	\includegraphics[width=\textwidth]{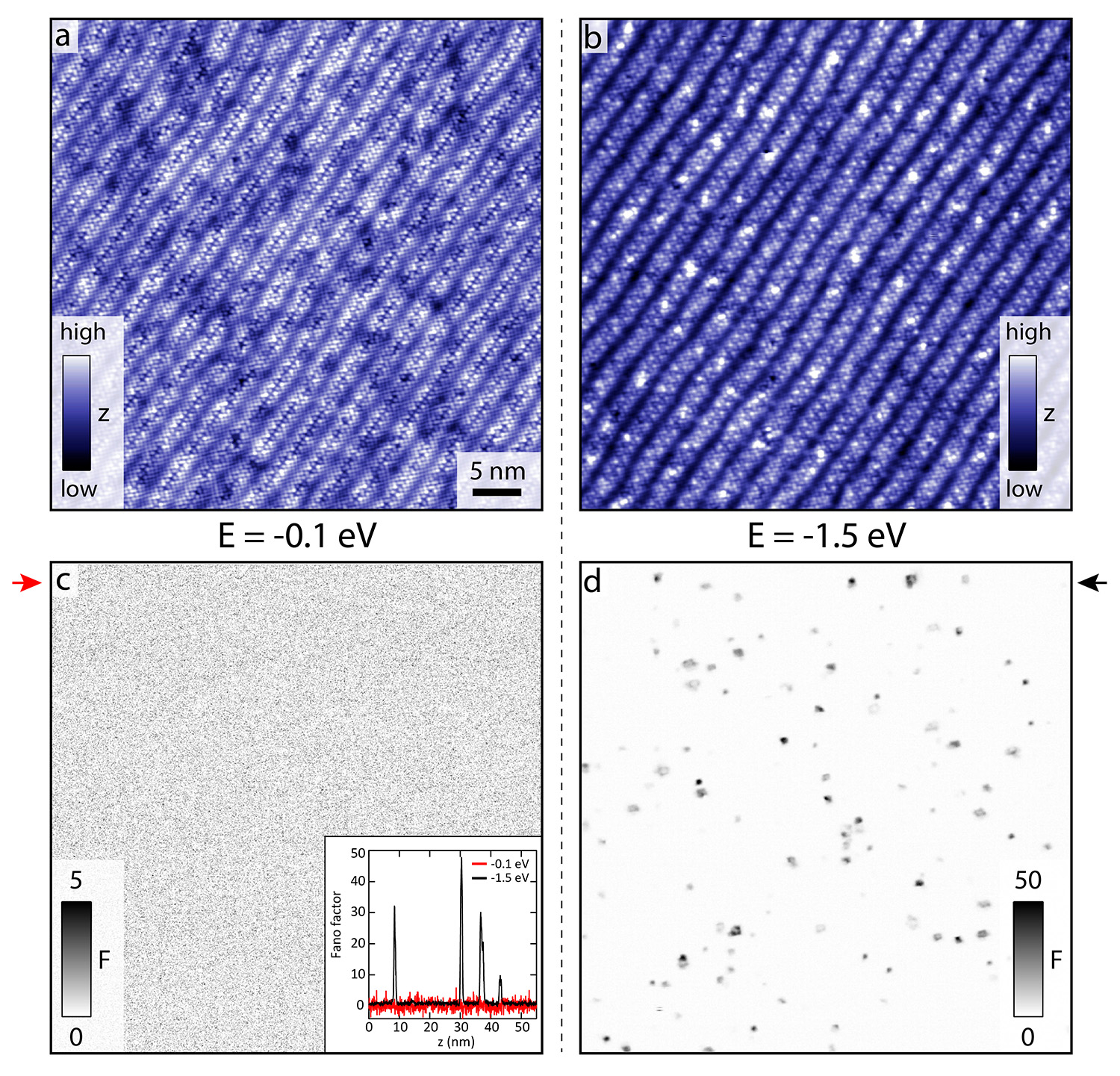}
	\caption{\label{fig:1} \textbf{Noisy defects in Bi2212}. \textbf{a} Constant current image (55 nm) of Bi2212, E = -100 meV, I = 100 pA. \textbf{b} Constant current image taken directly after \textbf{a} with E = -1.5 eV, I = 400 pA. \textbf{c, d} Shot-noise images simultaneously taken with \textbf{a}, \textbf{b} respectively. The inset in \textbf{c} shows a horizontal line cut through \textbf{c} (red) and \textbf{d} (black) at the position indicated by the arrows. Note: the bright spots in \textbf{b} do \textit{not} correspond to the locations with excess shot-noise in \textbf{d}, see also Supplementary Information section I.}
\end{figure}

A typical constant current image of cleaved Bi2212 is shown in Fig. \ref{fig:1}a, displaying the characteristic incommensurate super-modulation, and clear atomic contrast of the Bi atoms in the BiO surface termination plane. Numerous studies focussing on the tunnelling paths in Bi2212 have attributed this surface appearance to the insight that the main tunnelling path from the (conducting) copper oxygen plane below the surface through the (insulating) buffer layers to the tip is through the bismuth atoms directly above the copper atoms, whereas the oxygen atoms in the CuO plane cannot be resolved due to a $\pi$-phase difference \cite{martin_prl_2002, he_prl_2006, zhou_prl_2007, nieminen_prb_2009, nieminen_prl_2009, zeljkovic_nanoletters_2014}. Upon adding a dopant atom, the situation changes. In general, the interstitial oxygen atoms have a net negative charge from donating one or two holes to the CuO plane. For sufficiently high positive bias of the tip (negative bias of the sample) this charge can be removed. If the oxygen dopant is coupled strongly enough to the CuO charge reservoir, the removed electron will be replaced immediately, resulting in the opening of an additional conduction channel. Previous studies have identified two types of such oxygen dopant atoms (at E = -1\thinspace eV and E = -1.5\thinspace eV), as well as oxygen vacancies (at E = +1\thinspace eV), through the enhancement of the local density of states \cite{kinoda_prb_2003, kinoda_prb_2005, mcelroy_science_2005, zeljkovic_science_2012, zeljkovic_nanoletters_2014}. At the negative sample bias we focus on in this study, we find both types of oxygen dopant and even resolve the previously predicted anisotropic shape for those found at E = -1.5\thinspace eV \cite{zeljkovic_nanoletters_2014}, see Supplementary Information section I. The difference in appearance of the two types of oxygen atoms comes from the observation that although both dopants are below the BiO layer, those found at E = -1.5\thinspace eV are located in between two Bi atoms, whereas the E = -1\thinspace eV dopants lie in between a square of four Bi atoms \cite{zeljkovic_science_2012, zeljkovic_nanoletters_2014}. Other than the presence of these dopant states, the general tunnelling process from the CuO plane to the tip is unaffected: atomic contrast and the super-modulation are nearly identical at E = -1.5\thinspace eV to that at lower energy, see Fig. \ref{fig:1}b; the additional bright spots correspond to the dopant states that appear at this energy. 

In contrast to the dopant atoms discussed thus far that are strongly coupled to the charge reservoir, local environments are conceivable where this is not the case. Once an electron is removed from such a dopant, it cannot be replaced immediately, leading temporarily to a locally charged (or strictly speaking less charged) state. The charging and de-charging of such a dopant and the accompanying fluctuations in the local potential will completely change the tunnelling dynamics. In general, a strongly fluctuating local potential will have two immediate effects on tunnelling through the dominant current carrying conduction channel(s): the differential conductance will go down as the number of states available for tunnelling is intermittently reduced, and the current noise will increase as tunnelling will take place in bursts that mimic bunching. Additionally, the current noise, in analogy with the current dependence of the charging and de-charging of defects that leads to 1/f noise, will become quadratically instead of linearly dependent on the current \cite{vanderziel_1978}. In fact, in addition to the known (strongly coupled) oxygen dopants, we find defects with spectral features at different energies, a significant number of which display a clear suppression of the density of states. Moreover, unlike the Poissonian shot-noise (F = 1) we measure at low energies (see Fig. \ref{fig:1}c)\footnote{The signal-to-noise for this particular measurement is set to be able to resolve changes in the slope of a factor of $\sim$5. This can be greatly improved upon by longer averaging or point spectroscopy, see Ref. [\citen{massee_arxiv_2018}].}, at high energy the same atomically sized regions that show additional spectral features display a strongly enhanced current-noise (F $\gg$ 1), see Fig. \ref{fig:1}d. The slope of the shot-noise in these regions can be up to tens of times higher than for the random tunnelling seen at low energies.

\begin{figure}
	\centering
	\includegraphics[width=\textwidth]{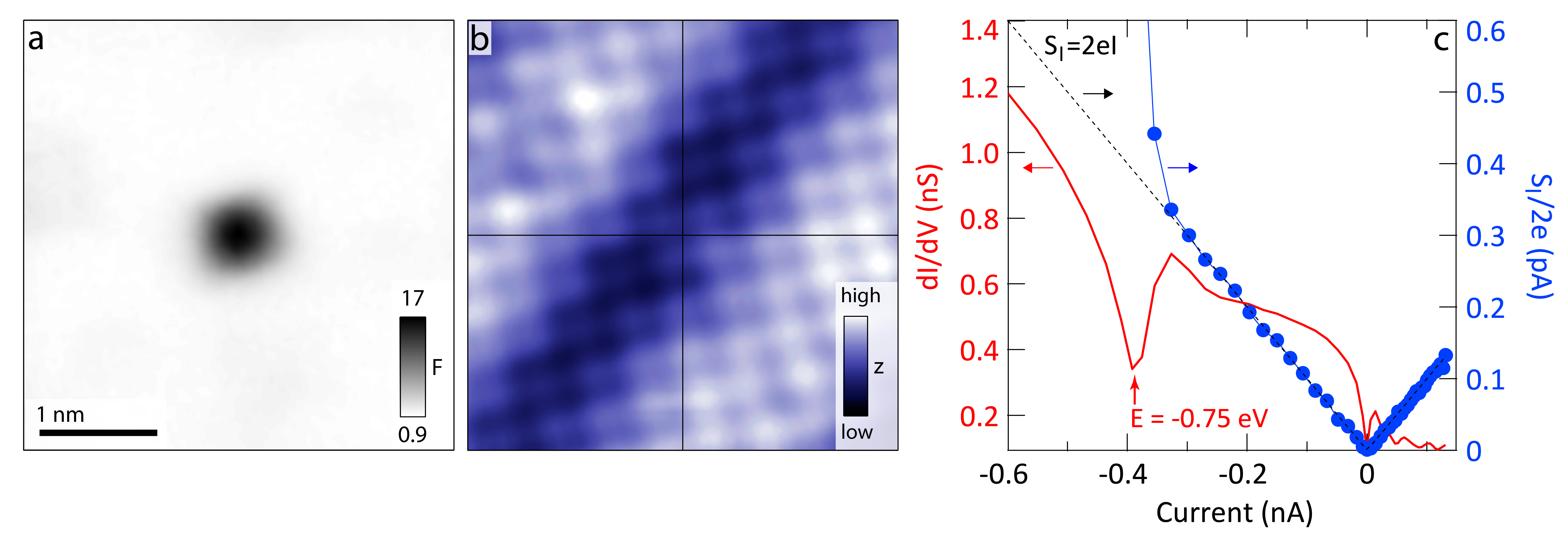}
	\caption{\label{fig:2} \textbf{General characteristics of noisy defects}. \textbf{a} Average shot-noise image constructed from 44 defects in Fig. \ref{fig:2}d. \textbf{b} Average topographic image corresponding to \textbf{a}. The clear atomic contrast indicates that the defects are all located at roughly the same location, i.e. in between four Bi atoms. \textbf{c} Typical differential conductance spectrum plotted versus the current (solid red, left axis, E$_{\text{setup}}$ = -1\thinspace eV, the dip is located at E = -0.75\thinspace eV) and simultaneously recorded noise (blue markers and line, right axis): the drop in differential conductance is strongly correlated with the deviation of the noise from the Poissonian value (dashed line).}
\end{figure}

Could the defects with excess, or super-Poissonian, shot-noise be charging and de-charging dopant atoms? To investigate them in more detail we first focus on their shape and spatial location. As the images in Fig. \ref{fig:1} show, the `noisy' defects are not located in the topmost BiO layer - there is no signature in topography indicating their presence. Furthermore, their predominant profile in the current noise is approximately spherical, although their appearance can be clover-like and energy dependent leading to more complicated shapes which we will discuss later. To determine the average shape of the defects and their spatial location, we extract the centre of each defect that is sufficiently isolated from neighbouring ones by fitting an xy-symmetric 2D Gaussian. For Fig. \ref{fig:1}d this amounts to 44 defects. We then define an equally sized window around the centre of each defect and construct the average Fano- and topographic images. Fig. \ref{fig:2}a shows the average defect seen through the shot-noise: an isotropic sphere with a diameter of $\sim$ 8\AA\ and a magnitude of F$\sim$17 (at E = -1.5\thinspace eV). If the defects had been situated at a random xy-location in the unit cell, the atomic contrast of the averaged topographic image shown in Fig. \ref{fig:2}b would have been washed out. Instead, we maintain clear atomic contrast, indicating that the defects are all located roughly at the same xy-location in the unit cell. This location, which corresponds to the centre of the images in Fig. \ref{fig:2}a-b, is in the middle of a square of Bi atoms - exactly like the oxygen dopant atoms found at -1\thinspace eV which have a comparable size and shape \cite{mcelroy_science_2005, zeljkovic_science_2012, zeljkovic_nanoletters_2014}, but no excess current noise.

To find out more, we turn our attention to spectroscopy. The spatially averaged spectrum of Bi2212 at negative energies is dominated by the appearance of the Cu-O and O bands at E $\lesssim$ -1\thinspace eV \cite{hybertsen_prl_1988, wells_prb_1989}, with localised peaks in the density of states due to single oxygen dopants \cite{kinoda_prb_2003, kinoda_prb_2005, mcelroy_science_2005, zeljkovic_science_2012, zeljkovic_nanoletters_2014}. In agreement with tunnelling affected by local charge modulations, the defects with super-Poissonian noise are generally characterised by a suppression of the differential conductance instead of a peak. This suppression is in most cases preceded by a modest increase at slightly lower energy, or occurs on the flank of a steep increase. A typical example of a spectrum taken on top of a `noisy' defect and the simultaneously recorded shot-noise is shown in Fig. \ref{fig:2}c. Clearly, the deviation from Poissonian noise (i.e. F = 1) is strongly correlated with the drop in differential conductance. 

\begin{figure}
	\centering
	\includegraphics[width=\textwidth]{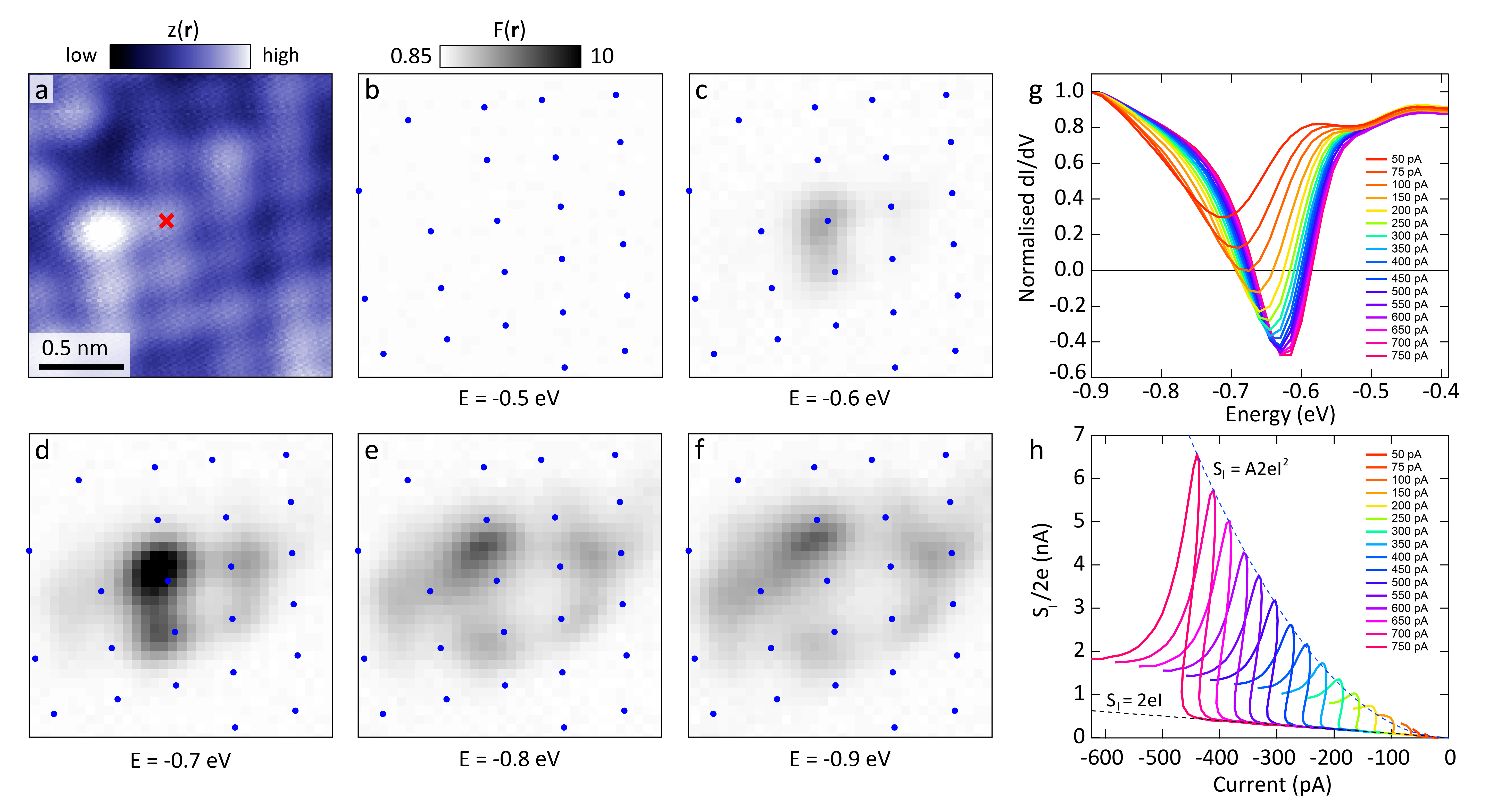}
	\caption{\label{fig:3} \textbf{Negative differential conductance and dispersion}. \textbf{a} Constant current image of an area with a noisy defect (E = -100\thinspace meV, I = 100\thinspace pA). \textbf{b-f} Spatially resolved shot-noise for five different energies: the noise initially appears (in \textbf{c}) on top of four neighbouring Bi atoms, but then radially moves outwards for increasing energies (each energy is recorded at I = 400\thinspace pA). Blue dots mark the location of the Bi atoms in \textbf{a}. \textbf{g} Differential conductance taken on the Bi atom marked in panel a) for increasing setup currents (E$_{\text{setup}}$ = -1\thinspace eV), becoming negative for I $>$ 100\thinspace pA. \textbf{h} Shot-noise taken simultaneously with \textbf{g}. Poissonian (S$_{I}$ = 2eI) and a quadratic dependence of the noise on current (S$_{I}$ = A2eI$^{2}$) are indicated with dotted lines.}
\end{figure}

Although most of the noisy defects we observe on Bi2212 show a modest decrease in differential conductance, a few actually display negative differential conductance (NDC). Instead of a point-like spatial appearance in the centre of a square of four Bi atoms as shown in Fig. \ref{fig:2}a, the enhanced noise of these defects is initially located on top of the four neighbouring Bi atoms and radially expands up to one or two lattice sites with increasing energy, see Fig. \ref{fig:3}a-f. Point spectra taken on the Bi atom indicated with a cross in Fig. \ref{fig:3}a at low currents display similar behaviour to that in Fig. \ref{fig:2}c: a slight increase in differential conductance followed by a drop. However, for increasing currents, the initial increase is suppressed and for I $>$ 100\thinspace pA the differential conductance becomes negative as can be seen in Fig. \ref{fig:3}g. The simultaneously recorded shot-noise has a pronounced peak coinciding with the drop in differential conductance. As expected for charge modulated current, the maximum of the shot-noise is quadratic in current (Fig. \ref{fig:3}h). We further note that even though the current-noise decreases after reaching a peak, it does not return to its Poissonian value, an observation that is true for all noisy defects we have observed.

\begin{figure}
	\centering
	\includegraphics[width=\textwidth]{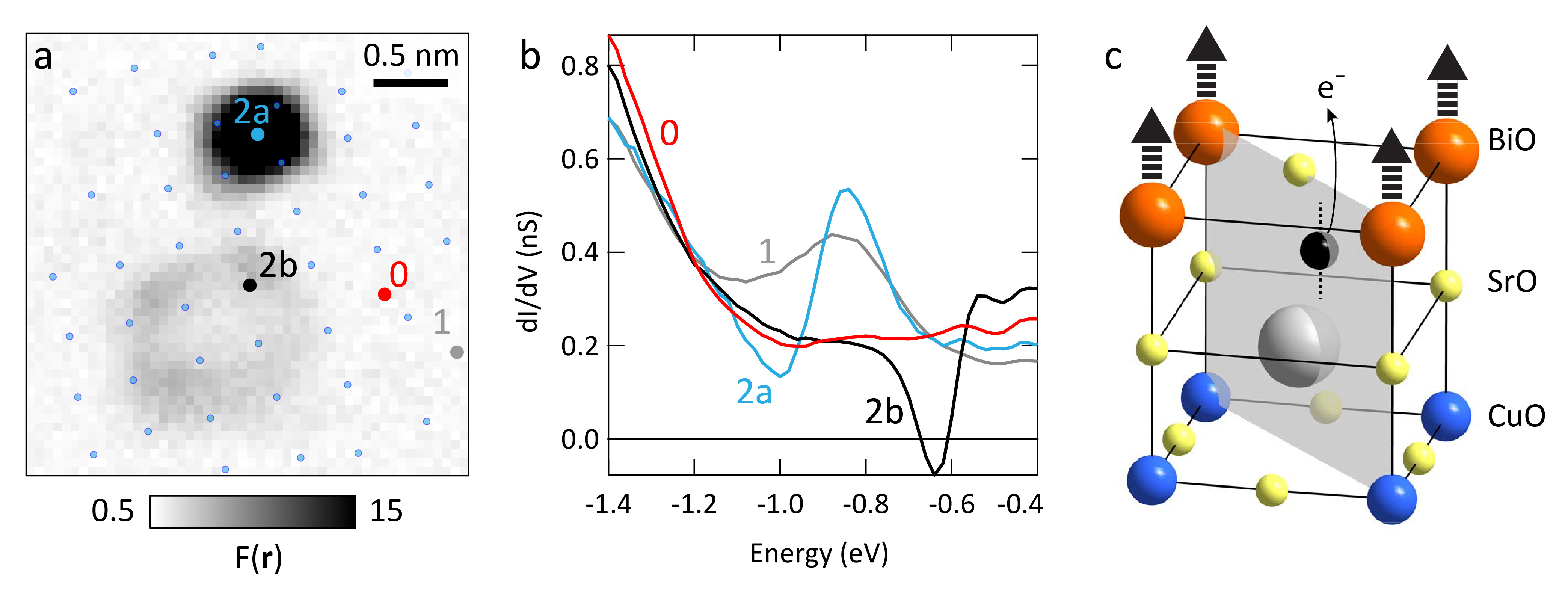}
	\caption{\label{fig:4} \textbf{All-in-one: different manifestations of defects on the same xy-location.} \textbf{a} Fano factor on a 3\thinspace nm field of view (E = -1.4\thinspace eV, I = 400\thinspace pA) with three different defects: one conventional (\#1) and two noisy (\#2a,b). A location without defect is also marked (\#0). Blue dots mark the location of the Bi atoms. \textbf{b} Differential conductance taken at the locations indicated in \textbf{a}. \textbf{c} Top three layers of cleaved Bi2212, the dopant location from Ref. [\citen{he_prl_2006}] in agreement with our data is shown in black (the dotted line represents possible height variations of the dopant). Weak coupling to the CuO plane of some defects due to their local environment leads to charging and de-charging that modulates the current through the main conduction channels - for the bottom defect in \textbf{a} this is through the four neighbouring Bi atoms.}
\end{figure}

Based on the atomic scale size and the location between four Bi atoms, a natural explanation for the super-Poissonian noise we observe in Bi2212 is that tunnelling is strongly affected by a charging and de-charging oxygen dopant atom. The earlier observation that there are too few conventional defects (at E = -1\thinspace eV and E = -1.5\thinspace eV) for the expected oxygen doping concentration \cite{mcelroy_science_2005, zeljkovic_science_2012} strongly favours this assignment. This implies, however, that certain oxygen dopants that would otherwise have lead to a conventional resonance at E = -1\thinspace eV, instead appear as noisy defects. Figure \ref{fig:4}a exemplifies this by showing a Fano image on a 3\thinspace nm field of view containing three defects we encounter on the same xy-position relative to the unit cell: (1) a conventional one with a resonance around E = -1\thinspace eV, (2a) a noisy one with the excess current noise appearing directly above the defect, and (2b) a noisy one with excess noise and negative differential conductance appearing on the four Bi atoms neighbouring the defect. The spectra corresponding to the three defects, as well as that of a region without one (0), are shown in Fig. \ref{fig:4}b. The atomic structure of the topmost three layers of cleaved Bi2212 and the lowest energy position of a dopant atom (black) \cite{he_prl_2006} in agreement with the location of the defects is shown in Figure \ref{fig:4}c.

The question is how oxygen dopant atoms, which are located at the same xy-location within experimental error, can manifest themselves so dramatically differently. One important clue comes from the current dependence of the defects. In semiconductor systems, a strong dependence of the differential conductance on the setup conditions (i.e. current) is known to be caused by the electric field from the tip that locally bends the bands up- or downwards at the surface, i.e. tip induced band bending (TIBB) \cite{feenstra_jvactechnolb_1987}. In essence, for increasing tunnelling currents at the same voltage, the tip is closer to the surface, which increases the tip induced electric field and reduces the barrier for tunnelling from dopants near the surface to the tip. Crucially, the effect is strongest for dopants closest to the surface. Clearly, the defect with negative resistance (Fig. \ref{fig:3}) displays a strong current dependence of the differential conductance and accompanying shot-noise. In general, nearly all noisy defects show a finite current dependence, in contrast to the conventional resonances that show no dependence on the setup current (see Supplementary Information section III). This observation strongly suggests that a key parameter for how a dopant atom reveals itself is its distance from the BiO plane. The distance, combined with the complicated crystal structure, possible nearby elemental substitutions, other defects and vacancies, and the non-commensurate super-modulation, then provides a range of local environments where the coupling of a dopant atom can become weak enough to lead to the array of observed atomic scale charge dynamics.

A consequence of the reduced tunnel barrier for dopants close to the surface due to TIBB is an increase in the rate of ionization. Therefore, for higher setup currents at the same voltage, the average time the dopant is positively (strictly speaking less negatively) charged due to an electron tunnelling to the tip will increase, reducing the differential conductance to lower - or even negative - values and increasing the effective bunching, all in agreement with our observations. Additionally, TIBB is known to produce an energy dependent halo around charged dopants in semiconductors like GaAs and InAs \cite{lee_nanolett_2004, teichmann_prl_2008, marczinowski_prb_2008}, where the halo reflects the energy dependent lateral distance from which a dopant can be ionised. The dispersive current noise in Fig. \ref{fig:3}b-f, which has a one-to-one correspondence to changes in the differential conductance (see Supplementary Information section II), is consistent with this mechanism. 

Although the general behaviour of the differential conductance - a dip preceded by an increase - and appearance of super-Poissonian noise is very reminiscent of resonant tunnelling in a double barrier structure \cite{brown_ieee_1992, jahan_sse_1995, iannaccone_prl_1998, prb_kuznetsov_1998, blanter_prb_1999}, there is one crucial difference: for resonant tunnelling the noise returns to its Poissonian value once the energy reaches the end of the resonance. This also holds for dynamical Coulomb blockaded tunnelling due to an interacting localised state \cite{safonov_prl_2003, thielmans_prb_2005, djuric_apl_2005, zarchin_prl_2007, kim_acsnano_2010, xue_scientificreports_2015}, where a quadratic dependence of the noise on the current has been observed \cite{safonov_prl_2003}. In contrast, the shot-noise of our noisy defects does not return to the Poissonian value of F = 1. The fact that we still observe excess current noise at E = -1.5\thinspace eV in Fig. \ref{fig:1}d despite all resonant energies being at lower energies is a direct visualization thereof. Unlike resonant tunnelling, charging and de-charging can still take place at higher energies, resulting in enhanced current noise even at energies above the resonant energy. The magnitude of the current noise then reflects the derivative of the differential conductance with respect to the fluctuating potential. As shown in Supplementary Information section IV, we have evidence that this is indeed the case.

Could our observations, instead of the result of charge dynamics at oxygen dopants, somehow be caused by an artefact of the tip, sample contamination, the AC circuitry we use to measure the shot-noise, or local heating? To exclude tip effects, we have used tungsten and Pt/Ir tips, both giving identical results. Furthermore, intentionally damaging either tip material has no effect other than reducing the spatial resolution (see Supplementary Information section V). Multiple as-grown, as well as annealed samples have been studied, all showing similar behaviour, ruling out sample contamination or unusual inhomogeneous samples as culprit. With our ability to completely disconnect our AC circuitry without changing tunnelling conditions, we confirmed that the defect in Fig. \ref{fig:3} displays negative differential conductance both with and without AC circuit (see Supplementary Information section V). Lastly, the noise power spectral density we measure is the sum of the shot-noise and the thermal noise. In principle, local (energy dependent) heating could therefore enhance the apparent shot-noise. However, to reach noise levels we observe on a number of defects would require local electron temperatures of $>$20\thinspace K, which is highly unlikely for our sample temperature of $<$2\thinspace K. 

To summarize, by combining scanning tunnelling microscopy and shot-noise measurements, we resolve atomic scale defects where the current noise is strongly enhanced and the differential conductance reduced - for some defects even negative. These noisy defects most likely constitute a subset of the oxygen dopant atoms that are introduced to turn the Mott insulator Bi2212 into a high temperature superconductor, dopants that were unaccounted for in previous scanning probe studies. The locally enhanced current noise and reduced differential conductance is expected for charging and de-charging on time scales of the tunnelling process due to weak coupling of these dopants to the CuO charge reservoir, resulting in local potential fluctuations that strongly affect the dominant tunnelling channel(s). Given the similar xy-location, the exact details of the local environment plays a key role in the charge dynamics of oxygen dopants and their effect on the tunnelling process. Their unconventional behaviour, in particular of the defects with negative differential conductance, can in principle be utilised to dynamically control the doping at the atomic scale, enabling direct visualization of the effect of local charging on e.g. high T$_{\text{c}}$ superconductivity. 

\begin{methods}
High quality Bi$_{2}$Sr$_{2}$CaCu$_{2}$O$_{8+x}$ single crystals were grown at the University of Amsterdam using the floating zone technique. Both as-grown samples with a T$_{c}$ $\sim$ 90\thinspace K and samples annealed for 5 days at 450\thinspace $^{\circ}$C under 20\thinspace mbar oxygen pressure with a T$_{c}$ $\sim$ 81\thinspace K were studied. The samples were mechanically cleaved in cryogenic vacuum at T $\sim$ 20\thinspace K and directly inserted into the STM head at 4.2 K. Etched atomically sharp and stable tungsten tips, and mechanically cut Pt/Ir tips were used, both with energy independent density of states. Differential conductance measurements throughout used a standard lock-in amplifier with a modulation frequency of 429.7Hz. STM and simultaneous noise measurements were performed with the home-built setup and MHz circuitry described in Ref. [\citen{massee_arxiv_2018}]. A bandpass filter followed by a Herotek DZM020BB diode was used to integrate the noise amplitude spectral density in the 100\thinspace kHz - 5\thinspace MHz frequency range. Lock-in measurements of the noise at the LC$_{\text{cable}}$ resonance of 1\thinspace MHz gave identical results. All presented measurements were recorded at T = 1.8\thinspace K.
\end{methods}

\begin{addendum}
 \item We thank J. C. Davis, A. Mesaros, D. K. Morr, D. Roditchev and P. Simon for fruitful discussions. FM would like to acknowledge funding from H2020 Marie Sk\l{}odowska-Curie Actions (grant number 659247) and the ANR (ANR-16-ACHN-0018-01).
 \item[Author contributions] F.M. and M.A. conceived the study. F.M. designed and built the microscope, performed and analysed all measurements and wrote the manuscript with valuable input from M.S.G. and M.A.. Samples were grown by Y.H.K..
\end{addendum}


\begin{thebibliography}{}	
	\bibitem{uren_apl_1985}
	M. J. Uren, D. J. Day and M. J. Kirton, Appl. Phys. Lett. \textbf{47}, 1195 (1985) \textit{1/f and random telegraph noise in silicon metal-oxide semiconductor field-effect transistors}
	
	\bibitem{chen_science_1999}
	J. Chen, M. A. Reed, A. M. Rawlett, J. M. Tour, Science \textbf{286}, 1550 (1999) \textit{Large on-off ratios and negative differential resistance in a molecular electronic device}
	
	\bibitem{banin_nature_1999}
	U. Banin, Y. Cao, D. Katz and O. Millo, Nature \textbf{400}, 542 (1999) \textit{Identifcation of atomic-like electronic states in indium arsenide nanocrystal	quantum dots}
	
	\bibitem{pradhan_prl_2005}
	N. A. Pradhan, N. Liu, C. Silien and W. Ho, PRL \textbf{94}, 076801 (2005) \textit{Atomic scale conductance induced by single impurity charging}
	
	\bibitem{loth_science_2012}
	S. Loth, S. Baumann, C. P. Lutz, D. M. Eigler, A. J. Heinrich, Science \textbf{335}, 196 (2012) \textit{Bistability in atomic-scale antiferromagnets}
	
	\bibitem{schaffert_naturematerials_2013}
	J. Schaffert, M. C. Cottin, A. Sonntag, H. Karacuban, C. A. Bobisch,
	N. Lorente, J. -P. Gauyacq and R. M\"{o}ller, Nature Materials \textbf{12}, 223 (2013) \textit{Imaging the dynamics of individually adsorbed molecules}
	
	\bibitem{lit_naturecommunications_2013}
	J. van der Lit, M. P. Boneschanscher, D. Vanmaekelbergh, M. Ij\"{a}s, A. Uppstu, M. Ervasti, A. Harju, P. Liljeroth and I. Swart, Nature Communications \textbf{4}, 2023 (2013) \textit{Suppression of electron–vibron coupling in	graphene nanoribbons contacted via a single atom}
	
	\bibitem{wickenburg_naturecommunications_2016}
	S. Wickenburg, J. Lu, J. Lischner, H. -Z. Tsai, A. A. Omrani, A. Riss, C. Karrasch, A. Bradley, H. Sae Jung, R. Khajeh, D. Wong, K. Watanabe, T. Taniguchi, A. Zettl, A. H. Castro Neto, S. G. Louie and M. F. Crommie, Nature Communications \textbf{7}, 13553 (2016) \textit{Tuning charge and correlation effects for a single molecule on a graphene device}
	
	\bibitem{repp_science_2004}
	J. Repp, G. Meyer, F. E. Olsson and M. Persson, Science \textbf{305}, 493 (2004) \textit{Controlling the charge state of individual gold atoms}
	
	\bibitem{lee_nanolett_2004}
	D. -H. Lee and J. A. Gupta, Nano Letters \textbf{11}, 2004-2007 (2004) \textit{Tunable control over the ionization state of single Mn acceptors in GaAs with defect-induced band bending}

	\bibitem{teichmann_prl_2008}
	K. Teichmann, M. Wenderoth, S. Loth, R. G. Ulbrich, J. K. Garleff, A. P. Wijnheijmer and P. M. Koenraad, Phys. Rev. Lett. \textbf{101}, 076103 (2008) \textit{Controlled charge switching on a single donor with a scanning tunneling microscope}

	\bibitem{marczinowski_prb_2008}
	F. Marczinowski, J. Wiebe, F. Meier, K. Hashimoto and R. Wiesendanger, Phys. Rev. B \textbf{77}, 115318 (2008) \textit{Effect of charge manipulation on scanning tunneling spectra of single Mn acceptors in InAs}

	\bibitem{rashidi_naturecommunications_2016}
	M. Rashidi, J. A. J. Burgess, M. Taucer, R. Achal, J. L. Pitters, S. Loth and R. A. Wolkow, Nature Communications \textbf{7}, 13258 (2016) \textit{Time-resolved single dopant charge dynamics in silicon}
	
	\bibitem{blanter_physicsreports_2000}
	Ya. M. Blanter and M. B\"{u}ttiker, Physics Reports \textbf{336}, 1-166 (2000) \textit{Shot noise in mesoscopic conductors}
	
	\bibitem{blanter_2010}
	Ya. M. Blanter \textit{Recent Advances in Studies of Current Noise} In: Vojta M., R\"{o}thig C., Sch\"{o}n G. (eds) CFN Lectures on Functional Nanostructures - Volume 2. Lecture Notes in Physics, vol 820. Springer, Berlin, Heidelberg
	
	\bibitem{massee_arxiv_2018}
	F. Massee, Q. Dong, A. Cavanna, Y. Jin and M. Aprili, arxiv:1806.00372 (2018) \textit{Atomic scale shot-noise using broadband scanning tunnelling microscopy}
	
	\bibitem{bastiaans_arxiv_2018}
	K. M. Bastiaans, T. Benschop, D. Chatzopoulos, D. H. Cho, Q. Dong, Y. Jin and M. P. Allan, arxiv:1806.00374 (2018) \textit{Amplifier for scanning tunneling microscopy at MHz frequencies}
	
	\bibitem{martin_prl_2002}
	I. Martin, A. V. Balatsky and J. Zaanen, Phys. Rev. Lett. \textbf{88}, 097003 (2002) \textit{Impurity states and interlayer tunneling in high temperature superconductors}

	\bibitem{he_prl_2006}
	Y. He, T. S. Nunner, P. J. Hirschfeld, and H. -P. Cheng, PRL \textbf{96}, 197002 (2006) \textit{Local electronic structure of Bi$_{2}$Sr$_{2}$CaCu$_{2}$O$_{8}$ near oxygen dopants, a window on the high-T$_{c}$ pairing mechanism}
	
	\bibitem{zhou_prl_2007}
	S. Zhou, H. Ding and Z. Wang, Phys. Rev. Lett. \textbf{98}, 076401 (2007) \textit{Correlating off-stoichiometric doping and nanoscale electronic inhomogeneity in the high-T$_{c}$ superconductor Bi$_{2}$Sr$_{2}$CaCu$_{2}$O$_{8+\delta}$}
	
	\bibitem{nieminen_prb_2009}
	J. Nieminen, I. Suominen, R. S. Markiewicz, H. Lin and A. Bansil, Phys. Rev. B \textbf{80}, 134509 (2009) \textit{Spectral decomposition and matrix element effects in scanning tunneling spectroscopy of Bi$_{2}$Sr$_{2}$CaCu$_{2}$O$_{8+\delta}$}
	
	\bibitem{nieminen_prl_2009}
	J. Nieminen, H. Lin, R. S. Markiewicz and A. Bansil, Phys. Rev. Lett. \textbf{102}, 037001 (2009) \textit{Origin of the electron-hole asymmetry in the scanning tunneling spectrum of the high-temperature Bi$_{2}$Sr$_{2}$CaCu$_{2}$O$_{8+\delta}$ superconductor}

	\bibitem{zeljkovic_nanoletters_2014}
	I. Zeljkovic, J. Nieminen, D. Huang, T. -R. Chang, Y. He, H. -T. Jeng, Z. Xu, J. Wen, G. Gu, H. Lin, R. S. Markiewicz, A. Bansil
	and J. E. Hoffman, Nano Letters \textbf{14}, 6749 (2014) \textit{Nanoscale interplay of strain and doping in a high-temperature superconductor}

	\bibitem{kinoda_prb_2003}
	G. Kinoda and T. Hasegawa, Phys. Rev. B \textbf{67}, 224509 (2003) \textit{Observations of electronic inhomogeneity in heavily pb-doped Bi$_{2}$Sr$_{2}$CaCu$_{2}$O$_{y}$ single crystals by scanning tunneling microscopy}
	
	\bibitem{kinoda_prb_2005}
	G. Kinoda, H. Mashima, K. Shimizu, J. Shimoyama, K. Kishio and T. Hasegawa, Phys. Rev. B \textbf{71}, 020502(R) (2005) \textit{Direct determination of localized impurity levels located in the blocking layers of Bi$_{2}$Sr$_{2}$CaCu$_{2}$O$_{y}$ using scanning tunneling microscopy/spectroscopy}
	
	\bibitem{mcelroy_science_2005}
	K. McElroy, J. Lee, J. A. Slezak, D. -H. Lee, H. Eisaki, S. Uchida and J. C. Davis, Science \textbf{309}, 1048 (2005) \textit{Atomic-scale sources and mechanism of nanoscale electronic disorder in Bi$_{2}$Sr$_{2}$CaCu$_{2}$O$_{8+\delta}$}
	
	\bibitem{zeljkovic_science_2012}
	I. Zeljkovic, Z. Xu, J. Wen, G. Gu, R. S. Markiewicz and J. E. Hoffman, Science \textbf{337}, 320 (2012) \textit{Imaging the impact of single oxygen atoms on superconducting Bi$_{2+y}$Sr$_{2-y}$CaCu$_{2}$O$_{8+x}$}

	\bibitem{vanderziel_1978}
	A. van der Ziel and E. R. Chenette, Advances in Electronics and Electron Physics \textbf{46}, 313–383 (1978) \textit{Noise in Solid State Devices}
	
	\bibitem{hybertsen_prl_1988}
	M. S. Hybertsen and L. F. Mattheiss, Phys. Rev. Lett. \textbf{60}, 1661 (1988) \textit{Electronic band structure of CaBi$_{2}$Sr$_{2}$Cu$_{2}$O$_{8}$}
	
	\bibitem{wells_prb_1989}
	B. O. Wells, P. A. P. Lindberg, Z. -X. Shen, S. D. Dessau, W. E. Spicer, I. Lindau, D. B. Mitzi and A. Kapitulnik, Phys. Rev. B \textbf{40}, 5259 (1989) \textit{Valence-band states in Bi$_{2}$(Ca,Sr,La)$_{3}$Cu$_{2}$O$_{8}$}
		
	\bibitem{feenstra_jvactechnolb_1987}
	R. M. Feenstra and J. A. Stroscio, J. Vac. Sci. Technol. B \textbf{5}, 923 (1987) \textit{Tunneling spectroscopy of the GaAs(110) surface}

	\bibitem{brown_ieee_1992}
	E. R. Brown, IEEE Transactions on Electron Devices \textbf{39}, 2686 (1992) \textit{Analytic model of shot noise in double-barrier resonant-tunneling diodes}
	
	\bibitem{jahan_sse_1995}
	M. M. Jahan and A. F. M. Anwar, Solid-State Electronics \textbf{38}, 429 (1995) \textit{Shot noise in double barrier quantum structures} 
	
	\bibitem{iannaccone_prl_1998}
	G. Iannaccone, G. Lombardi, M. Macucci and B. Pellegrini, Phys. Rev. Lett. \textbf{80}, 1054 (1998) \textit{Enhanced shot noise in resonant tunneling: theory and experiment}
	
	\bibitem{prb_kuznetsov_1998}
	V. V. Kuznetsov, E. E. Mendez, J. D. Bruno and J. T. Pham, Phys. Rev. B \textbf{58}, 10159(R) (1998) \textit{Shot noise enhancement in resonant-tunneling structures in a magnetic field}
	
	\bibitem{blanter_prb_1999}
	Ya. M. Blanter and M. B\"{u}ttiker, Phys. Rev. B \textbf{59}, 10217 (1999) \textit{Transition from sub-Poissonian to super-Poissonian shot noise in resonant quantum wells}
	
	\bibitem{safonov_prl_2003}
	S. S. Safonov, A. K. Savchenko, D. A. Bagrets, O. N. Jouravlev, Y. V. Nazarov, E. H. Linfield and D. A. Ritchie, Phys. Rev. Lett. \textbf{91}, 136801 (2003) \textit{Enhanced shot noise in resonant tunneling via interacting localized states}
	
	\bibitem{thielmans_prb_2005}
	A. Thielmans, M. H. Hettler, J. K\"{o}nig and G. Sch\"{o}n, Phys. Rev. B \textbf{71}, 045341 (2005) \textit{Super-Poissonian noise, negative differential conductance, and relaxation effects in transport through molecules, quantum dots, and nanotubes}
	
	\bibitem{djuric_apl_2005}
	I. Djuric, B. Dong and H. L. Cui, Appl. Phys. Lett. \textbf{87}, 032105 (2005) \textit{Super-Poissonian shot noise in the resonant tunneling due to coupling with a localized level}
	
	\bibitem{zarchin_prl_2007}
	O. Zarchin, Y. C. Chung, M. Heiblum, D. Rohrlich and V. Umansky, Phys. Rev. Lett. \textbf{98}, 066801 (2007) \textit{Electron bunching in transport through quantum dots in a high magnetic field}
	
	\bibitem{kim_acsnano_2010}
	Y. Kim , H. Song, D. Kim, T. Lee and H. Jeong, ACS Nano \textbf{4}, 4426 (2010) \textit{Noise characteristics of charge tunneling via localised states in metal-molecule-metal junctions}
	
	\bibitem{xue_scientificreports_2015}
	H. -B. Xue, J. -Q. Liang and W. -M. Liu, Scientific Reports \textbf{5}, 8730 (2015) \textit{Negative differential conductance and super-Poissonian shot noise in single-molecule magnet junctions}
\end{thebibliography}
\end{document}